\documentclass[sigconf]{acmart}
\usepackage{todonotes}
\usepackage{subfigure}  
\usepackage{graphicx}    
\usepackage{caption}
\usepackage{hyperref}
\usepackage{algorithm}    
\usepackage{algpseudocode}  
\usepackage{multirow}
\usepackage{balance}
\AtBeginDocument{%
  \providecommand\BibTeX{{%
    \normalfont B\kern-0.5em{\scshape i\kern-0.25em b}\kern-0.8em\TeX}}}

\setcopyright{acmcopyright}
\copyrightyear{2020}
\acmYear{2020}

\acmConference[DYNAMICS 2020]{DYNAMICS 2020: DYnamic and Novel Advances in Machine Learning and Intelligent Cyber Security (DYNAMICS) Workshop}{Dec. 07, 2020}{Texas, USA}
\acmBooktitle{DYNAMICS 2020: DYnamic and Novel Advances in Machine Learning and Intelligent Cyber Security (DYNAMICS) Workshop, Dec. 07, 2020, Texas, USA}
\acmPrice{15.00}

\begin{document}

\title{Program Behavior Analysis and Clustering using Performance Counters}

\author{Sai Praveen Kadiyala}
\affiliation{%
  \institution{I$^2$R, A*STAR, Singapore}
}
\email{saipk@i2r.a-star.edu.sg}

\author{Akella Kartheek}
\authornote{This work has been done when the author was with I$^2$R,
  A*STAR, Singapore.}
\affiliation{%
  \institution{BITS-Pilani Hyderabad, India}
}
\email{sukruthkartheek@gmail.com}

\author{Tram Truong-Huu}
\affiliation{%
  \institution{I$^2$R, A*STAR, Singapore}
  }
\email{truonght@i2r.a-star.edu.sg}

\renewcommand{\shortauthors}{Praveen Kadiyala, et al.}

\begin{abstract}
  Understanding the dynamic behavior of computer programs during normal
  working conditions is an important task, which has multiple security
  benefits such as the development of behavior-based anomaly
  detection, vulnerability discovery, and patching. Existing works
  achieved this goal by collecting and analyzing various data
  including network traffic, system calls, instruction traces, etc. In
  this paper, we explore the use of a new type of data, performance
  counters, to analyze the dynamic behavior of programs. Using
  existing primitives, we develop a tool named \texttt{perfextract} to
  capture data from different performance counters for a program
  during its startup time, thus forming multiple time series to
  represent the dynamic behavior of the program. We analyze the
  collected data and develop a semi-supervised clustering algorithm that allows us to
  classify each program using its performance counter time series into
  a specific group and to identify the intrinsic behavior of that
  group. We carry out extensive experiments with $18$ real-world
  programs that belong to $4$ groups including web browsers, text
  editors, image viewers, and audio players. The experimental results
  show that the examined programs can be accurately differentiated
  based on their performance counter data regardless of whether programs are run in physical or virtual environments.  
\end{abstract}  


\maketitle

\section{Introduction} 
\label{sec:intro}

The latest trends in program analysis show an increase in emphasis on the
dynamic approach compared to the static
approach~\cite{or2019dynamic}. In the dynamic approach, programs are
executed in a controlled environment and their behavior is observed during the execution time. The
interaction of programs with the operating system (OS) and generated data
during the execution are collected and analyzed. High-level features
such as application programming interface (API) calls, network traffic
and registry changes have demonstrated the
effectiveness~\cite{salehi2017maar,zhang2020dynamic,das2015semantics,mathew2018api}. However,
programs are increasingly sophisticated, especially malicious programs
(malware), which usually exhibit evasive behavior when being run
in a controlled execution environment. This motivates the research
community to explore the use of low-level features such as performance
counters as they are more resistant to external
attacks~\cite{dinakarrao2019adversarial,harris2019cyclone,tahir2019browsers}. Performance
counters have been initially introduced for the purpose of analyzing
complex systems for debugging related issues. With the help of
available digital logic that increments the counters after the
occurrence of a specific event, these counters keep track of various
micro-architectural events, thus enabling dynamic monitoring of
program behavior. We believe that performance counter-based profilers
give deeper insights into system dynamics with lower overhead compared
to their software profiler counterparts. Nevertheless, performance
counter data also has its own drawback related to the non-determinism
characteristic: several performance counters generate different values at different execution times~\cite{das2019sok}. This makes the program analysis task based on performance counters more challenging, thus requiring a more robust approach. 

Advocating the use of performance counters, we carry out in this work
an empirical study on program behavior analysis using a machine
learning technique. We propose a data collection approach that could
overcome the drawback of performance counter data discussed
above. Precisely, rather than collecting performance counters data
just at only one time instant and use this data as the sole input for
analysis, we propose to collect performance counters of a program over a
long period (e.g., during startup time of the program), thus forming a
time series for each performance counter. We note that while there
exist a large number of performance counters, we focus on per-program
performance counters, which count the number of micro-architectural
events triggered by a specific program. We use a machine learning
technique to learn not only the correlation between performance
counters but also the change in the pattern of a specific performance
counter over time. This approach allows us simultaneously to represent
the spatial-temporal correlation and to identify the intrinsic dynamic
behavior of each program through its performance counter time
series. While collecting performance counter data is an easy task,
labeling data to use in supervised learning techniques is not
straightforward due to the involvement of large man-hours and the
requirement of specific domain knowledge. Thus, we adopt
a clustering algorithm, which is a semi-supervised learning technique,
to classify programs based on their performance counter time series
into different groups, each having different dynamic behavior. The
contribution of our work is summarized as follows. 
 
\begin{itemize}
    \item We develop a tool named \texttt{perfextract} using existing
      primitives (\texttt{TypePerf}\footnote{TypePerf:~\url{https://ss64.com/nt/typeperf.html}}) to collect data from various
      performance counters during a specific time window, e.g., during
      startup time of programs. We note that while \texttt{TypePerf}
      has been developed to collect data from a performance counter at
      a given time instant, \texttt{perfextract} allows us to
      simultaneously collect data from various performance counters
      and form time series for each counter. This allows us to capture
      the behavior of each performance counter over time. We also note
      that there exist $28$ per-program performance counters in
      Windows operating systems, we collect data from $23$ counters since the remaining $5$ counters are not useful for our work.  
    
    \item We use a semi-supervised clustering algorithm, seeded $k$-means
      clustering, to classify programs, each being represented by
      multiple time series of performance counters, into different
      groups. We also adopt seeded $k$-means clustering to develop a new
      algorithm that detects whether a new program has been installed
      or deviated from its ``normal'' behavior. We analyze the dynamic behavior of each program group based on the changes in each performance counter over time.
    
    \item We carry out extensive experiments with $18$ real-world Windows programs belonging to $4$ groups: web browsers, text editors, image viewers, and audio players. We run these programs in both physical and virtual environments and use \texttt{perfextract} to collect data before doing analysis.
\end{itemize}

The rest of paper is organized as follows. In Section~\ref{sec:rel_work}, we discuss related work on dynamic analysis of programs using both high-level and low-level features. In Section~\ref{sec:PADM}, we describe our data collection framework and dynamic behavior analysis technique. In Section~\ref{sec:exp_res}, we present the experiments with various scenarios and analysis of results before we conclude the paper in Section~\ref{sec:conclusion}.

\section{Related Work} 
\label{sec:rel_work}

Characterizing the behavior of programs based on their dynamic
features is an established methodology in the field of system
security. In~\cite{salehi2017maar}, Salehi \textit{et al.}  used API
call return value recorded during the runtime of a program as features in modeling the program behavior. In a novel fashion, Zhaoqi \textit{et al.} used API calls along with their arguments for analyzing program behavior~\cite{zhang2020dynamic}. In~\cite{das2015semantics}, Das \textit{et al.} developed a semantics-based online malware detection approach based on API call sequences. In~\cite{mathew2018api}, an RNN-LSTM model has been developed for program analysis. The authors used the top-ranked API call sequences based on their Term Frequency and Inverse Document Frequency (TF-IDF). 

There also exist several works that used low-level features for
program behavior analysis. In~\cite{dinakarrao2019adversarial},
Dinkarrao \textit{et al.} used performance counter data to craft
adversarial attacks on malware detectors. The authors focused on
malware in Linux operating systems (OS), which has less attraction
compared to Windows OS.  In~\cite{harris2019cyclone}, Harris
\textit{et al.} used the performance counter data for understanding
the cyclic interference caused by cache contention, thereby detecting
external attacks. In~\cite{tahir2019browsers}, the authors proposed to
use performance counter data for profiling of web browsers during
their runtime and detecting obfuscation codes hidden in web
browsers. The authors demonstrated the effectiveness of the proposed
approach with Linux OS. In~ \cite{patel2017analyzing}, Nisarg
\textit{et al.} used performance counters along with machine learning
models for malware detection in Linux OS. The authors used a
statistical-based methodology to select the performance counters that
are to be monitored. Similarly, in~\cite{ozsoy2016hardware}, Ozosoy
\textit{et al.} developed a malware-aware processor (MAP), which has a
dedicated component to detect malware based on performance
counters. In~\cite{tang2014unsupervised}, Tang \textit{et al.} used
performance counter data to differentiate the behavior of two Windows
applications: \texttt{Internet Explorer} and \texttt{Adobe Acrobat
  Reader}. On one hand, most of the above works
(except~\cite{tang2014unsupervised}) focused on Linux OS. On the other
hand, they did not consider the changes in performance counters over
time. In this work, we consider programs running in Windows OS and
develop an approach that considers the spatial-temporal correlation of
performance counter data in the form of time series. This may overcome
the effect of non-determinism of performance counter data.

\section{Program Behavior Analysis and Clustering Framework} 
\label{sec:PADM}

In this section, we present the proposed framework for program behavior analysis and clustering. We start with an overview description and then describe the details of framework implementation and algorithm.

\subsection{Framework Overview} 
\label{subsec:pm_overview}

\begin{figure*}[t]
    \centering
    \includegraphics[width=0.8\textwidth]{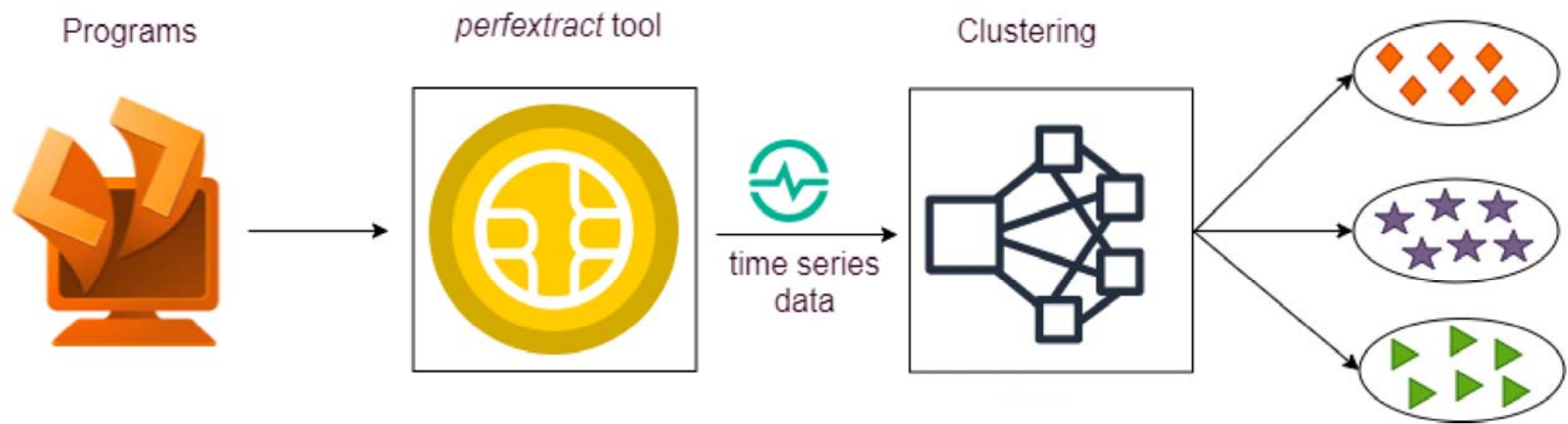}
    \caption{Framework overview.}
    \label{fig:proposed_flow}
    \vspace{-1.5ex}
\end{figure*}

In Fig.~\ref{fig:proposed_flow}, we depict the proposed framework, which consists of two components: data collection and program clustering. In the data collection component, we build up a program execution environment (e.g., a Windows virtual machine running in Oracle VirtualBox) with the \texttt{perfextract} tool integrated. This allows us to quickly refresh the environment after completing the analysis of each program. Given a program to be analyzed, we detonate the program in the execution environment and trigger the \texttt{perfextract} tool to collect performance counter data. The output of the data collection component is a set of time series that will be analyzed by a clustering algorithm.     

We advocate for clustering algorithms to classify programs into different groups since they do not need a labeled dataset, which could be hard to obtain in practice. We adopt $k$-means clustering to classify each program based on its time series of performance counters into different groups. We assume that there would be a normal working period that all the programs exhibit their normal behavior. The collected data during this normal working period creates the seeding clusters (e.g., the three clusters as shown on the right side of Fig.~\ref{fig:proposed_flow} and allows us to process new samples during the inference phase). Given a new data sample, the clustering algorithm will identify the best cluster that the sample belongs to. If the distance to the nearest cluster is larger than a threshold, the sample will form a new cluster, indicating that the program has not been recognized by the clustering algorithm. This can be due to two reasons: either the program has been newly installed in the system and it was run for the first time or the program has changed its behavior due to updates or security attacks.    

\vspace{-2ex}
\subsection{Extraction of Performance Counter Data}
\label{subsec:perfextracttool}

We now present the details of the extraction of performance counter data using the \texttt{perfextract} tool. We note that the number of performance counters could vary depending on the operating system run on the analyzing host. In Windows OS, there is a total of $28$ per-program performance counters that can be collected. However, during our experiments, we realized that there are $5$ counters whose values do not change over time. We decided to not collect the data from those counters as they do not contribute to differentiating the programs. The description of the remaining $23$ performance counters is given in Table~\ref{tab:des_perf_count}. 

\begin{table*}[t]
 \caption{Description of Per-program Performance Counters} 
  \label{tab:des_perf_count}
  \centering
 \begin{tabular}{cll}
 \hline
 \hline
\textbf{No.} & \textbf{Performance Counter} & \textbf{Description}  \\ \hline
1 & \%Privileged Time & Percentage
                                                          of elapsed
                                                          time that
                                                          the program
                                                          spent for executing code in privileged mode\\ \hline
2 & Handle Count & The total number of handles currently opened \\ \hline
3 & IO Read Operations/sec & The rate of read operations issued by the program \\ \hline
4 & IO Write Operations/sec & The rate of write operations issued by the program\\ \hline
5 & IO Data Operations/sec & The rate of read and write IO operations\\ \hline
6 & IO Other Operations/sec & The rate of IO operations other than read and write operations issued by the program\\ \hline
7 & IO Read Bytes/sec & The byte rate of read IO operations \\ \hline
8 & IO Write Bytes/sec & The byte rate write IO operations \\ \hline
9 & IO Data Bytes/sec & The byte rate of read and write IO operations\\ \hline
10 & IO Other Bytes/sec & The byte rate of IO operations other than read and write issued by the program \\ \hline
11 & Page Faults/sec & The rate of page faults occurred during its execution \\ \hline
12 & Page File Bytes Peak & Maximum number of bytes used in the paging file(s)\\ \hline
13 & Page File Bytes & Current number of bytes used in the paging file(s)\\ \hline
14 & Pool Paged Bytes & Current number of bytes in the paged pool\\ \hline
15 & Pool Non-paged Bytes & Current number of bytes in the non-paged pool\\ \hline
16 & Private Bytes & Current number of bytes that are not shared with other programs\\ \hline
17 & Priority Base &  Current base priority of the program\\ \hline
18 & Thread Count & Number of threads currently active in the program\\ \hline
19 & Virtual Bytes Peak & Maximum number of bytes of virtual address space used by the program\\ \hline
20 & Virtual Bytes & Current number of  bytes of the virtual address space used by the program \\ \hline
21 & Working Set Peak & Maximum number of bytes in the set of memory pages touched by the program \\ \hline
22 & Working Set & Current number of bytes in the set of memory pages touched by the program \\ \hline
23 & Working Set - Private & Current number of bytes in the set of memory pages exclusively touched by the programs \\ \hline
\end{tabular}
\end{table*}

For each performance counter, \texttt{perfextract} periodically
collects its value using a primitive available for Windows OS, i.e.,
\texttt{TypePerf}. The time interval between collection instants is
predefined (e.g., every $1$ second) so as to not cause significant
overhead (i.e., query the values too frequently) or miss the behavior
changes of the counter (i.e., the interval is too long). We also
define a time window for the collection duration so that the performance
counters of each program will be collected in a specific duration,
e.g., $30$ seconds. It is worth mentioning that it may be useful to
collect the data for the entire execution duration of the program as
the more data collected, the better the representation of program
behavior. However, the execution duration of each program depends on
the nature of its application and usage behavior of users, leading to
heterogeneous execution duration of programs. For instance, a user may
use a web browser for a few minutes for entertainment but he may use a
text editor for hours to work. Nevertheless, if the time window is too
short we may not have sufficient data for analyzing program
behavior. Furthermore, to avoid the bias from the usage behavior of users,
we collect the performance counter data during the start
time of programs, e.g., during the first $30$ seconds when programs
are launched. In actual implementation, we can use a program monitor
to trigger \texttt{perfextract} when a program is started. Collection
of performance counter data at the start time of programs may work
well with benign programs that do not exhibit evasive behavior. Thus,
there is a need for a more advanced approach (e.g., using a random
collection) to avoid such an evasive behavior. We keep this task for
future work when we work with malware samples. 

Consequently, we obtain $23$ time series, each having multiple values
(e.g., $30$ values collected over $30$ seconds) representing the
temporal evolution of a performance counter. By analyzing the time
series, we analyze the temporal behavior of performance counters
rather than comparing a particular value of performance counter among
programs. While two programs may have the same temporal behavior of
certain performance counters, they may differ from each other by the time series from the other performance counters. With the support of machine learning that could
learn the correlation from multi-dimensional data with both temporal and spatial correlation, we use performance counter time series to cluster programs into different groups, each having similar behavior. As discussed in the introduction, this could help one to identify unauthorized programs installed in the systems or detect abnormal behavior of a particular program, which has been previously installed and run in the system.

\subsection{Program Clustering and Fingerprinting}
\label{subsec:fingerprinting_method}

In this section, we present an algorithm that analyzes the data
collected from performance counters as presented in the previous
section to perform program clustering and fingerprinting. By
clustering, we mean that computer programs are classified into
different groups based on their functionalities tied to the
performance counter values, e.g., text editors,
web browsers, audio players, and image viewers. We refer to
this clustering approach as coarse-grained clustering. By
fingerprinting, we perform fine-grained clustering of data samples
(i.e., each data sample is a set of time series) collected at
different time instants of a particular program so as to identify
the program by its performance counter data. Obviously, the number of
groups (clusters) in the fine-grained clustering approach is much
larger compared to that of the coarse-grained approach. Given the
number of program groups and the total number of programs installed in
the system, we use a conventional clustering algorithm such as
$k$-means clustering to process the data. We note that $k$-means
clustering algorithm aims at minimizing the total distance of every
data sample to its cluster centroid. Given a dataset to be classified
into $k$ groups where $k$ is provided as an input, the partition of
the samples in the dataset is achieved by minimizing the following
objective function:
\begin{equation}
  \displaystyle\sum_{i=1}^k\sum_{x\in \mathcal{G}_i}||x - \mu_i||^2
 \end{equation}  
where $\mu_i$ is the centroid of cluster $\mathcal{G}_i$. Seeded
$k$-means~\cite{basu:2002} is a semi-supervised
clustering algorithm based on $k$-means. It used a small set of
labeled data to initialize $k$-means in computing the centroids rather
than choosing a random number of clusters. 

\begin{algorithm}[t]
\caption{\texttt{clusterProgram}($\mathcal{X}$)}
\label{alg:clustering}
\begin{algorithmic}[1]
\Require $\mathcal{X}$ \Comment{time series of performance counter data}
\Require $\mathcal{D}$ \Comment{existing labeled data samples}
\Ensure $\mathcal{C}$ \Comment{Cluster of the input sample}
\State $\{\mathcal{G},d\} \leftarrow \texttt{getNearestCluster}(\mathcal{X}, \mathcal{D})$\label{line:getnearestcluster}
\If {$d < \beta$} \Comment{$\beta$ is a predefined threshold}
\State Add sample $\mathcal{X}$ to cluster $\mathcal{G}$
\Else
\State Create a new cluster
\EndIf
\end{algorithmic}
\end{algorithm}
 
In Algorithm~\ref{alg:clustering}, we present the adopted clustering
algorithm for program clustering. Given a data sample that includes
$23$ time series of $23$ respective performance counters, the
algorithm starts by identifying the nearest cluster for the sample
based on the distance from the sample to the centroid of the
cluster. This step is performed by line~\ref{line:getnearestcluster}
in Algorithm~\ref{alg:clustering}. The output of the function
\texttt{getNearestCluster}($\mathcal{X}, \mathcal{D}$) is the index of
the nearest cluster ($\mathcal{G}$), and the distance between the
sample and the centroid of the cluster ($d$), normalized based on the
samples in the nearest cluster. If the distance is smaller than a predefined threshold
($\beta$), the sample is considered part of the cluster. Otherwise, it
will form a new cluster, which indicates that a new program has been
executed or a previously-seen program has deviated from its normal
behavior. It is to be noted that the function
\texttt{getNearestCluster}($\mathcal{X}, \mathcal{D}$) will compute
the distance from sample $\mathcal{X}$ to all the seeding
clusters. There may exist multiple seeding clusters that are close to
sample $\mathcal{X}$ with a distance smaller than $\beta$. The function
will return the closest cluster to the sample. The challenging issue
is how to determine the threshold to decide whether a program is
newly-installed in the system or it deviates from its previous behavior. A naive approach is to set
$\beta=1$, corresponding to the distance from the centroid to the farthest data sample
of the nearest cluster. However, this approach may not work well in
practice due to the non-linear characteristic of the multi-modal data
and the clusters are not in a spherical shape. There exist in the
literature other approaches that can be used to compute the distance
from a samples to a cluster centroid such as Euclidean distance or
$z$-score~\cite{thangavelu2018deft}. We note that Algorithm~\ref{alg:clustering} can be used for both cases: coarse-grained clustering and fine-grained clustering with a minor change in the parameter $k$ of the seeded $k$-means clustering algorithm.

\section{Experimental Results}
\label{sec:exp_res}

In this section, we present various experiments carried out to
demonstrate the effectiveness of the proposed approach. We first
present the experimental setup and the programs used in our
experiments before we present the analysis of results. 

\subsection{Experimental Setup and Programs}\label{subsec:setupdet}

\begin{table}[t]
\centering
\caption{Programs Used in  Experiments}\label{tab:dataset}
\begin{tabular}{|c|l|l|c|}
\hline
\textbf{No.} &  \textbf{Domain} & \textbf{Programs} & \textbf{\#Samples}\\ 
\hline
\multirow{2}{*}{1} & \multirow{2}{*}{Text editor} & Atom, gVim, Notepad,  & \multirow{2}{*}{20 each}\\
& & Notepad++, Sublime Text &\\
\hline
\multirow{3}{*}{2} & \multirow{3}{*}{Web browser} & Brave, Google Chrome, & \multirow{3}{*}{20 each}\\ 
& & Internet Explorer, Vivaldi, & \\
& & Firefox &\\
\hline
\multirow{3}{*}{3} & \multirow{3}{*}{Image viewer} & WildBit, ImageGlass, & \multirow{3}{*}{20 each}\\
& & FastStone, MS Paint,& \\
& & XnView &  \\
\hline
\multirow{2}{*}{4} & \multirow{2}{*}{Audio player} & Clementine, & \multirow{2}{*}{20 each} \\
& & foobar2000, VLC &   \\
\hline
\end{tabular}
\end{table}

\begin{figure}[t]
\centering
    \subfigure[Time series of Handle Counts]{\includegraphics[width=0.38\textwidth]{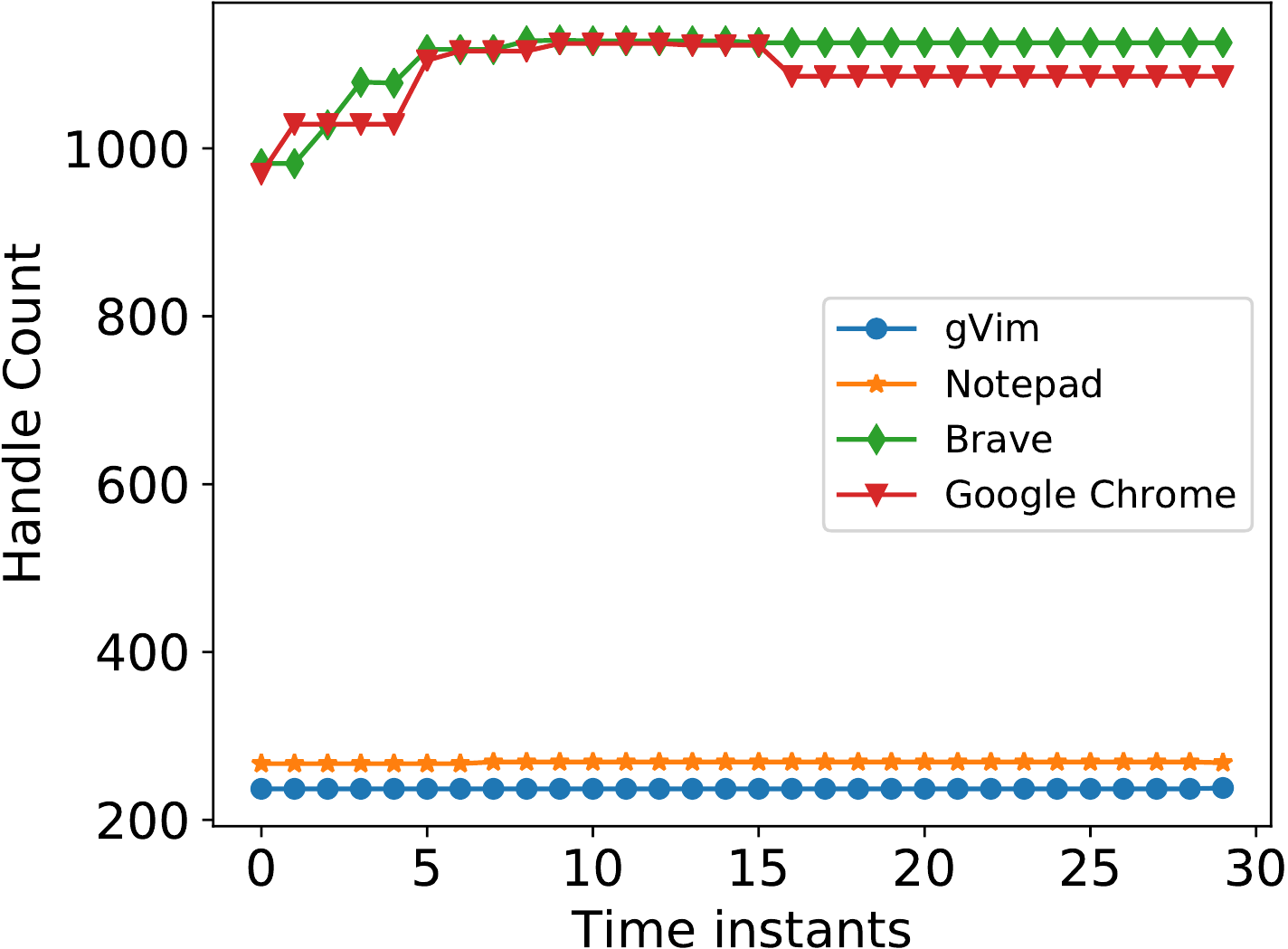}}
    \subfigure[Time series of page file bytes]{\includegraphics[width=0.38\textwidth]{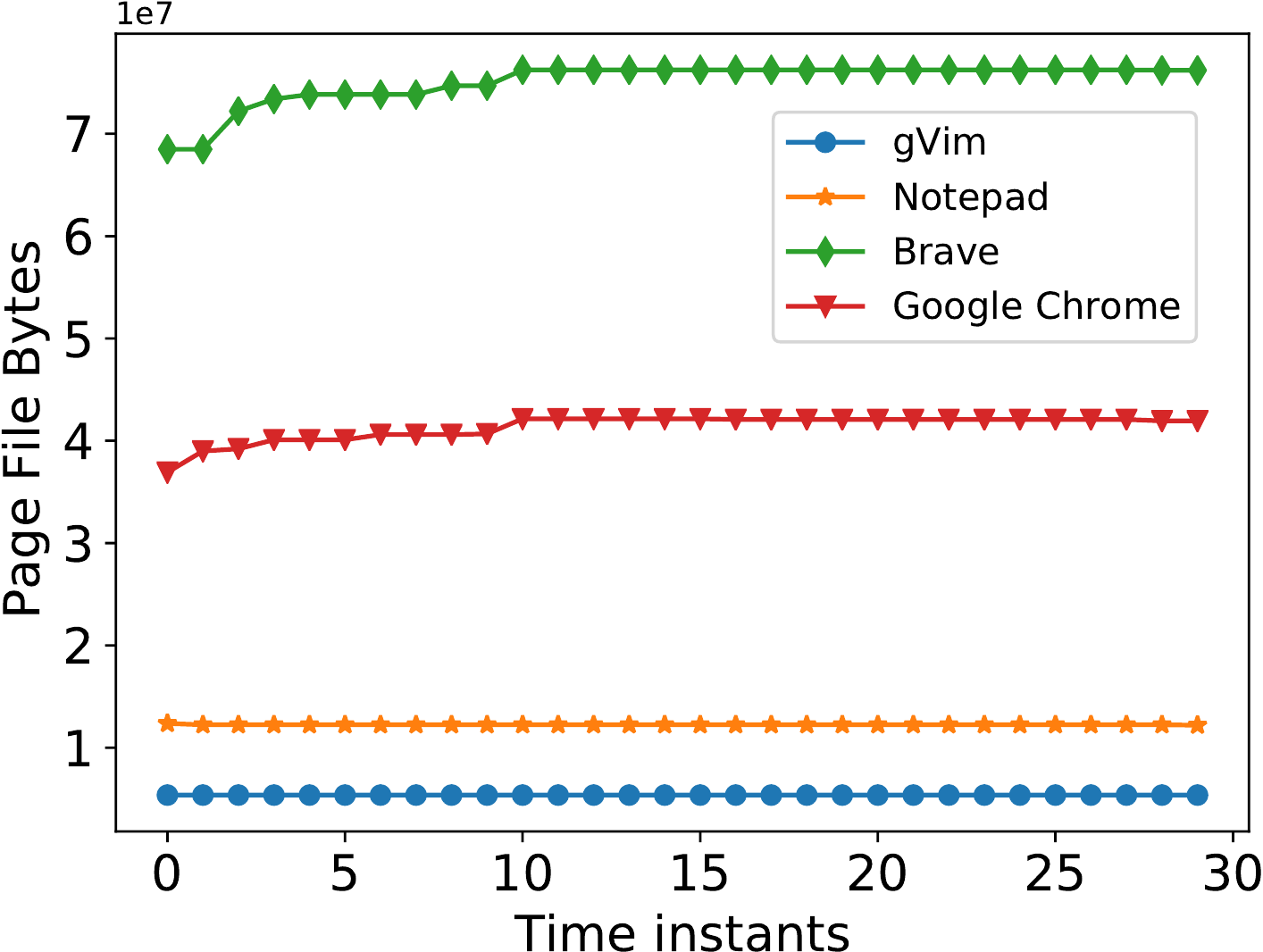}}
   	\caption{Time series of performance counter data of various programs.}
	\label{fig:time_series_browsers_text_editors}
        \vspace{-2ex}
      \end{figure}
      
We carried out all the experiments in a virtual machine with 16GB RAM
and Windows 10 Operating System using Oracle VirtualBox. We
implemented \texttt{perfextract} in Python, which also provides
various libraries for data processing and analysis. We used $18$
popular programs belonging to $4$ application domains including text
editors, web browsers, image viewers, and audio players. The selected
programs are presented in Table~\ref{tab:dataset}. For each of the
programs, we collected $20$ data samples by launching the program with
$20$ different input arguments, e.g., opening a web browser with $20$
randomly-chosen websites or launching a text editor with $20$
different text files. It is worth mentioning again that each data
sample includes $23$ time series of performance counters presented in
Table~\ref{tab:des_perf_count}, each having $30$ values for a duration
of $30$ seconds. Furthermore, to reduce the bias of the data to any
particular run, for each input argument, we executed the monitored
program (e.g., text editor, web browser) ten
times, each commencing at a random instance in time. This increases
the randomness of data so that the machine learning algorithms will
learn the data pattern (i.e., the temporal behavior of programs)
rather than a particular value. Consequently, we have collected $3600$
($18 \times 20 \times 10$) data samples for $18$ programs, which will
be used for the experiments. In
Fig.~\ref{fig:time_series_browsers_text_editors}, we plot the time
series of two performance counters (i.e., \texttt{Handle Counts} and \texttt{Page File
Bytes}) of various programs. We see that there is a significant
separation in the behavior of the performance counter data among
programs. This allows machine learning algorithms to learn program
behavior and differentiate them efficiently.

\subsection{Analysis of Results}

\subsubsection{Coarse-grained and Fine-grained Clustering.}

\begin{figure}[t]
\centering
\subfigure[Coarse-grained intra-cluster distance.]{\includegraphics[width=0.39\textwidth]{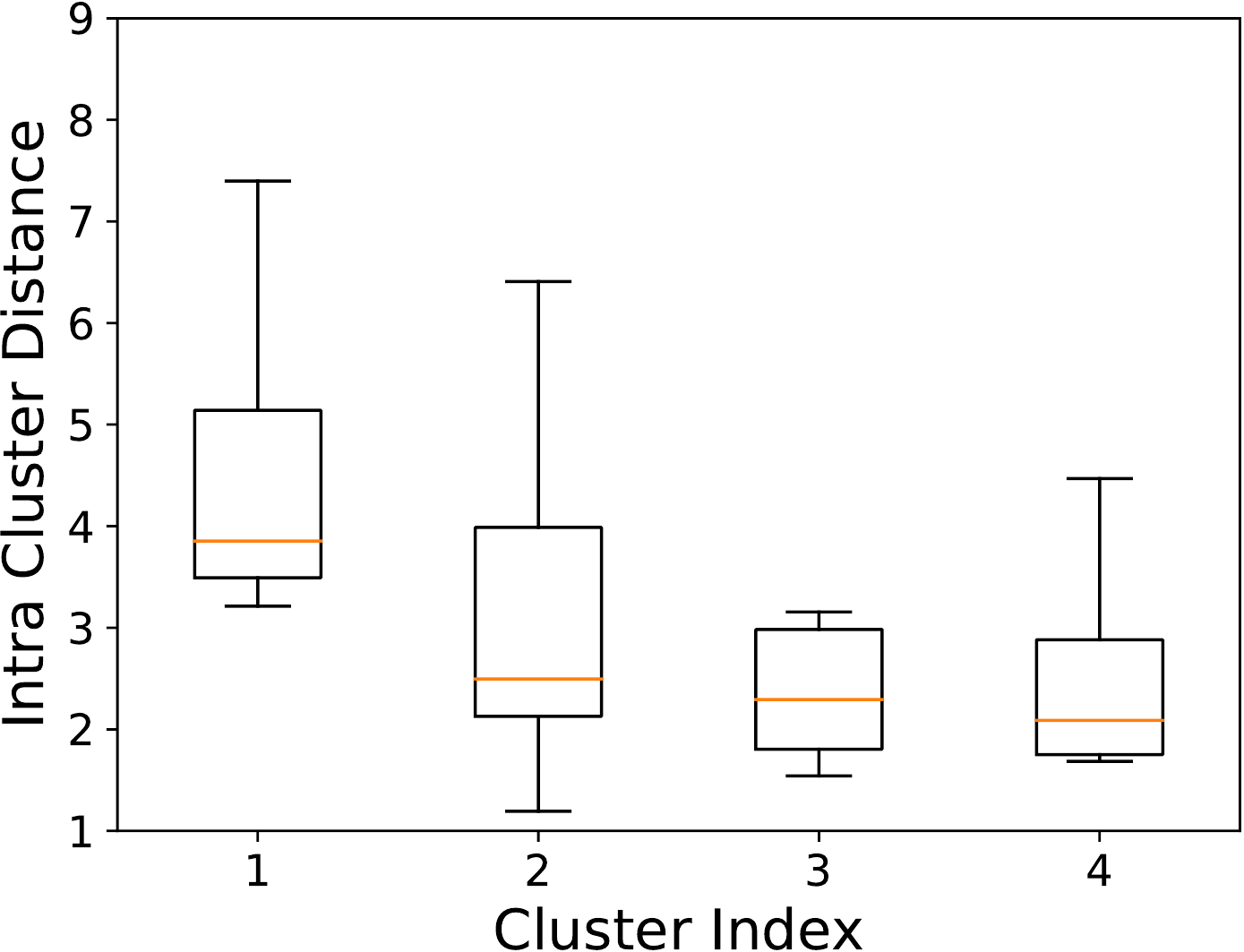}}
\subfigure[Coarse-grained inter-cluster distance.]{\includegraphics[width=0.39\textwidth]{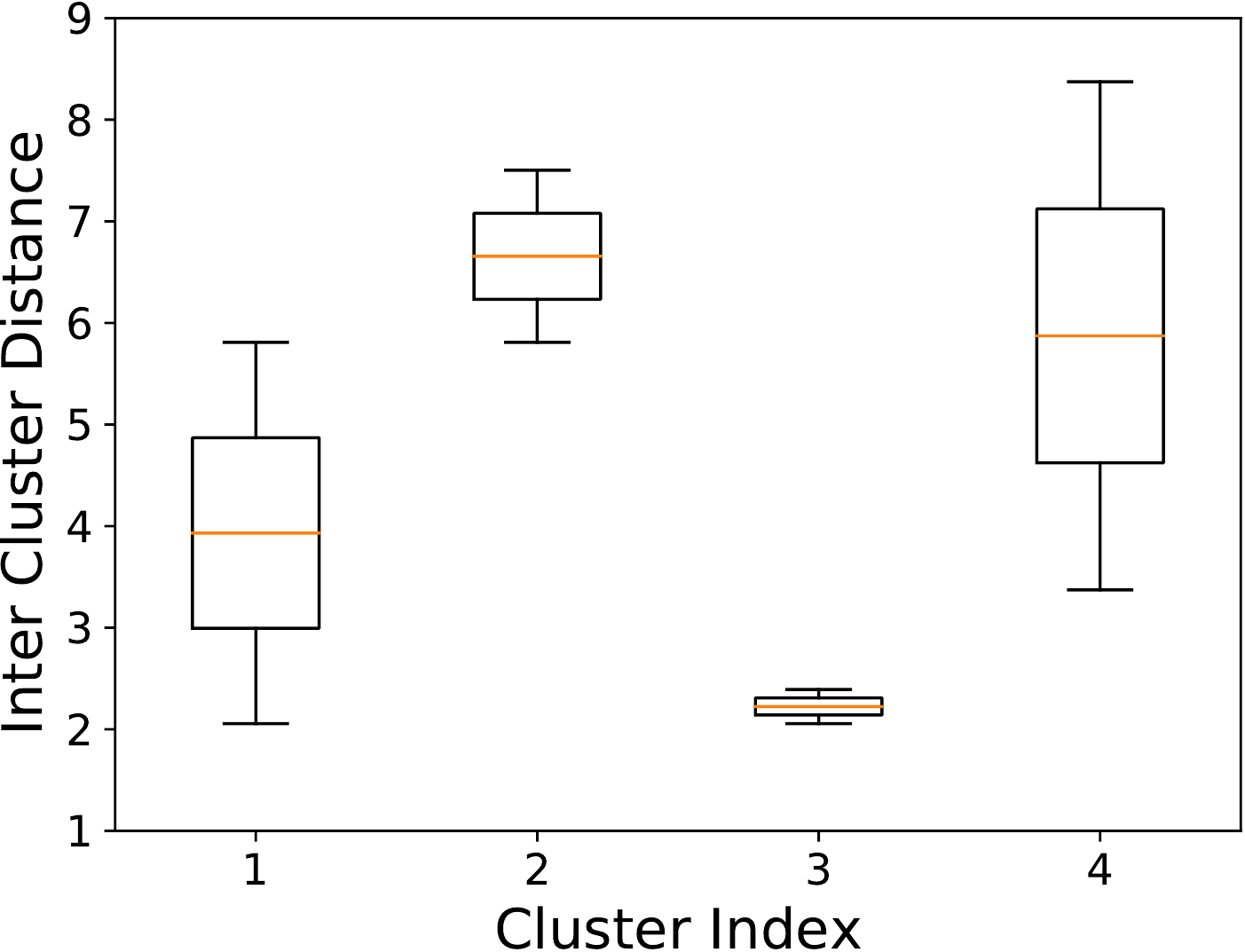}}
\caption{Box plot representation for intra-cluster and inter-cluster distances with coarse-grained approach.} 
\label{fig:inter_intra_cluster_dist_coarse}
\vspace{-1.5ex}
\end{figure}

\begin{figure}[t]
\centering
\subfigure[Fine-grained intra-cluster distance.]{\includegraphics[width=0.39\textwidth]{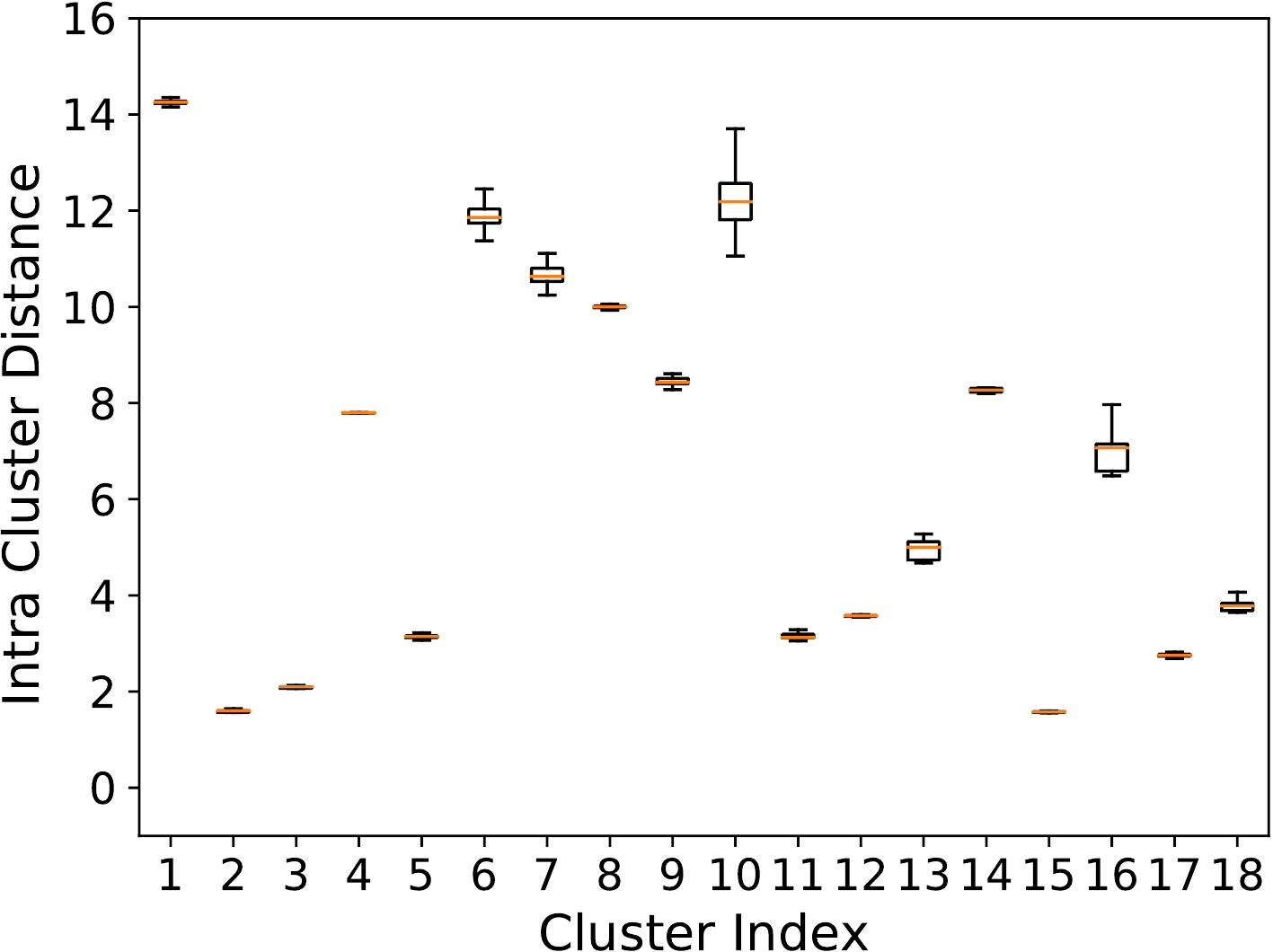}}
\subfigure[Fine-grained inter-cluster distance.]{\includegraphics[width=0.39\textwidth]{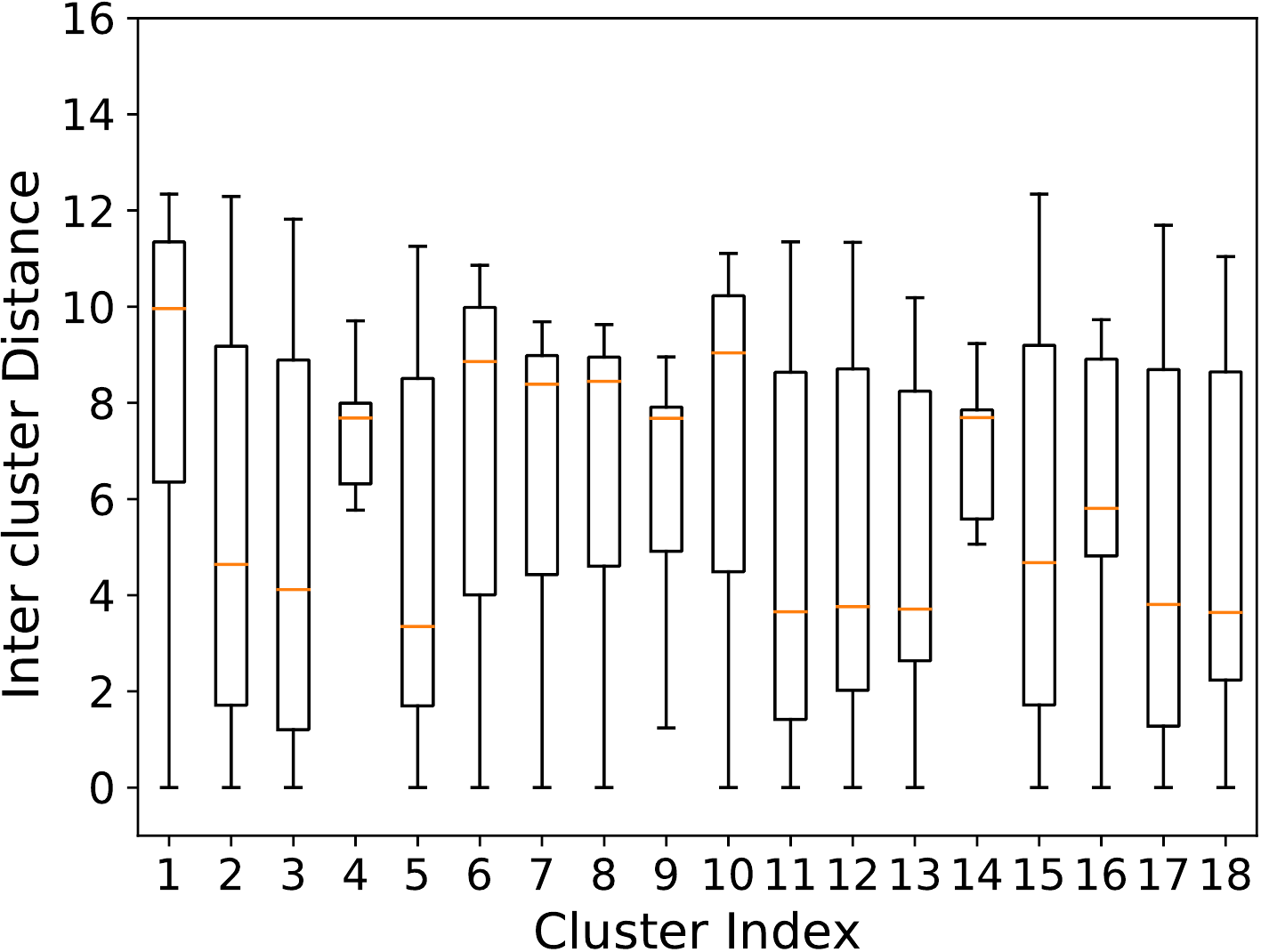}}
\caption{Box plot representation for intra-cluster and inter-cluster
  distances with fine-grained
  approach.} \label{fig:inter_intra_cluster_dist_fine}
\vspace{-1.5ex}
\end{figure}

In this experiment, we fed all $3600$ data samples to the clustering
algorithm ($k$-mean clustering), given the number of clusters (i.e., $4$ clusters in case
of coarse-grained clustering and $18$ clusters in case of fine-grained
clustering). In Fig.~\ref{fig:inter_intra_cluster_dist_coarse} and
Fig.~\ref{fig:inter_intra_cluster_dist_fine}, we present the box plot
representation of intra-cluster and inter-cluster distances for the
clusters resulted from the clustering algorithm. The intra-cluster
distance of a cluster is computed as the mean of the distance from all
the samples belonging to the cluster to its centroid. The
inter-cluster distance of a cluster is computed as the mean of the 
distance from its centroid to the centroid of the remaining
clusters. As expected, the intra-cluster distance of most of the clusters
is smaller than the inter-cluster distance. We also obtained a very
small standard deviation of the distances with the fine-grained
clustering scenario. This demonstrates that data samples belonging to the same cluster have similar behavior that allows the clustering algorithm to learn their similarity and pattern.

\begin{figure}[t]
    \centering
    \subfigure[Coarse-grained clustering]{\includegraphics[width=0.45\textwidth]{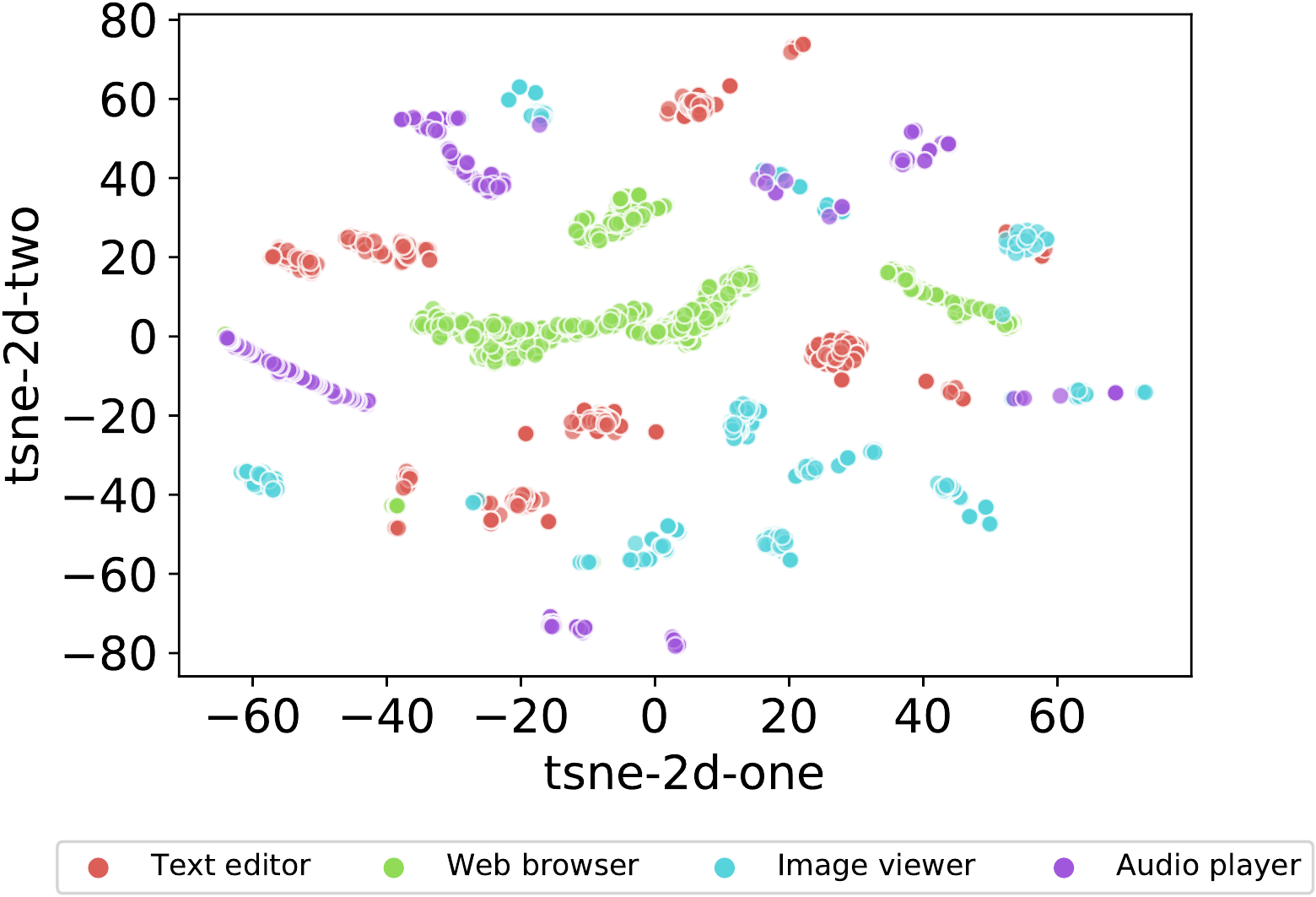}}
    \subfigure[Fine-grained clustering]{\includegraphics[width=0.45\textwidth]{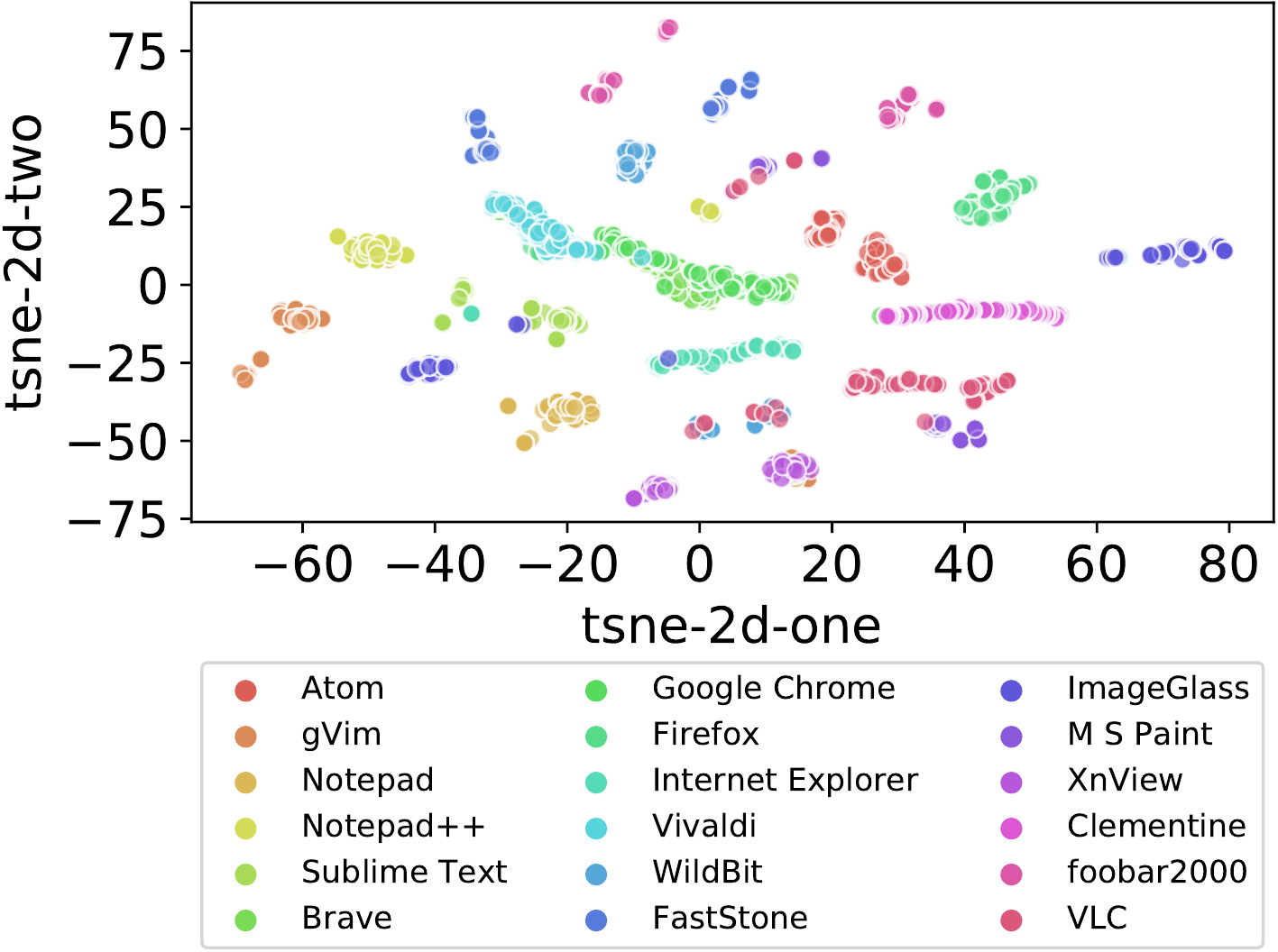}}
    \caption{t-SNE two-dimensional plots for results obtained by clustering.}
    \label{fig:coarse_vs_fine}
\end{figure}

To further demonstrate the effectiveness of our clustering approach,
we use t-SNE~\cite{hinton:2008} to produce the two-dimensional scatter
plots  of the clustering results using coarse-grained and fine-grained
approaches. We note that t-SNE is a technique for
visualization of high-dimensional data by applying dimensionality
reduction to the data before plotting to a two or three-dimensional
plot. In Fig.~\ref{fig:coarse_vs_fine}, the results show that even
though each group of programs is distributed at different locations,
there is a clear separation among clusters of programs and application
types. This insight provides us the motivation to develop the algorithm to detect new programs installed in the system or behavior deviation of existing programs.

\subsubsection{Identification of Unknown Programs}
\label{subsec:clus_res}

In this experiment, we evaluate the capability of the proposed
approach to detect unknown programs. Given the $18$ programs used for
our experiments, we
randomly selected $4$ programs, among them $3$ programs are considered
as seeding clusters and the remaining program is considered as an unknown
program. With seeding clusters, we mean that all the data samples of
the $3$ programs are analyzed by the clustering algorithm forming $3$
different clusters whose labels are known a priori. Given a data
sample of the unknown program, we used Algorithm~\ref{alg:clustering}
to compute the distance between the data sample and the centroids of
the $3$ seeding clusters and determine whether a new cluster is formed
or not. If a new cluster is formed, it indicates that the proposed
approach is able to detect the unknown program executed in the
system. Otherwise, it fails to do so.  

\begin{figure}[t]
\centering
\subfigure[Separation among Atom, Brave, WildBit and Clementine]{\includegraphics[width=0.45\textwidth]{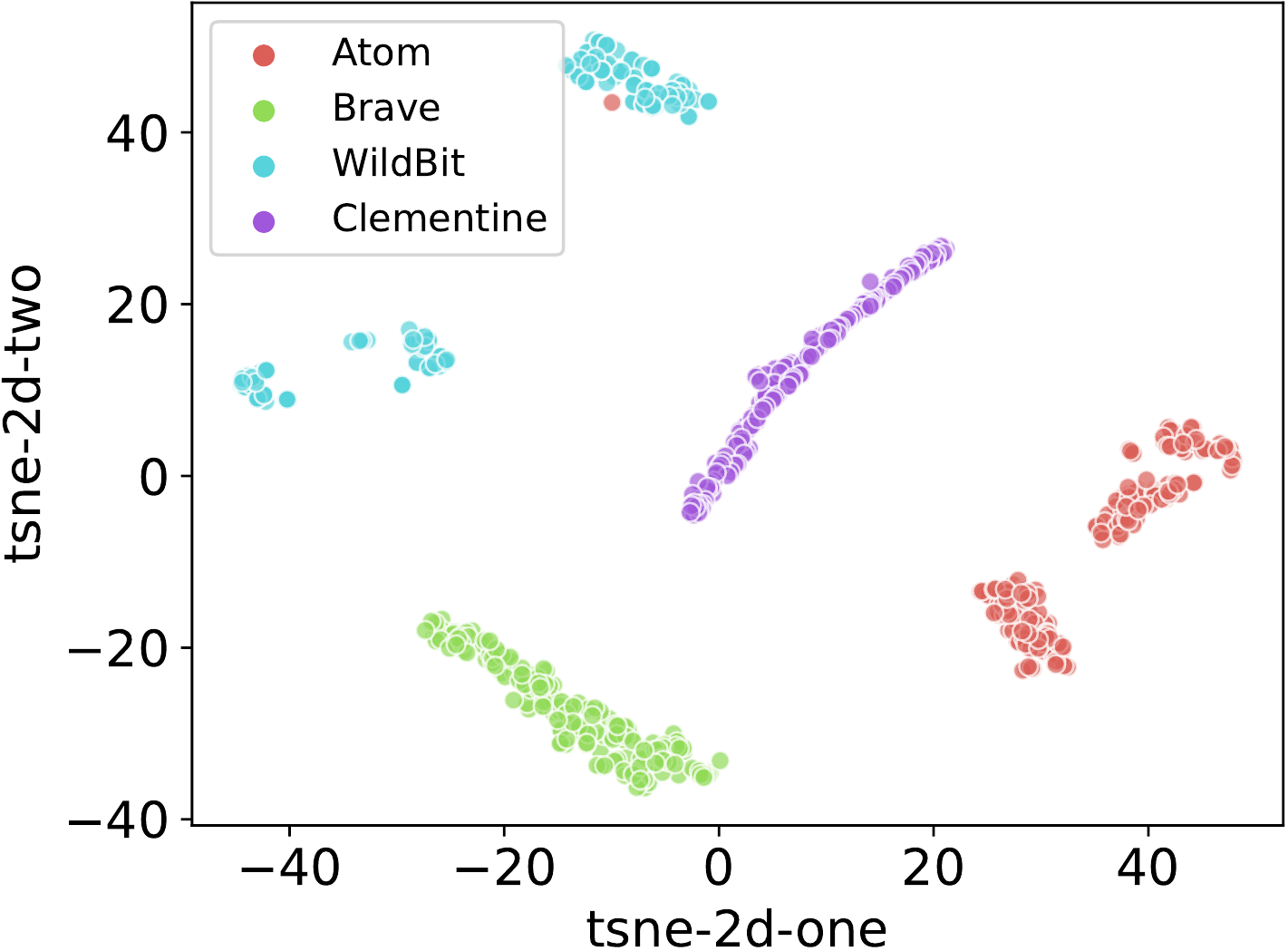}}
\subfigure[Separation among Vivaldi, Clementine, foobar2000 and VLC]{\includegraphics[width=0.45\textwidth]{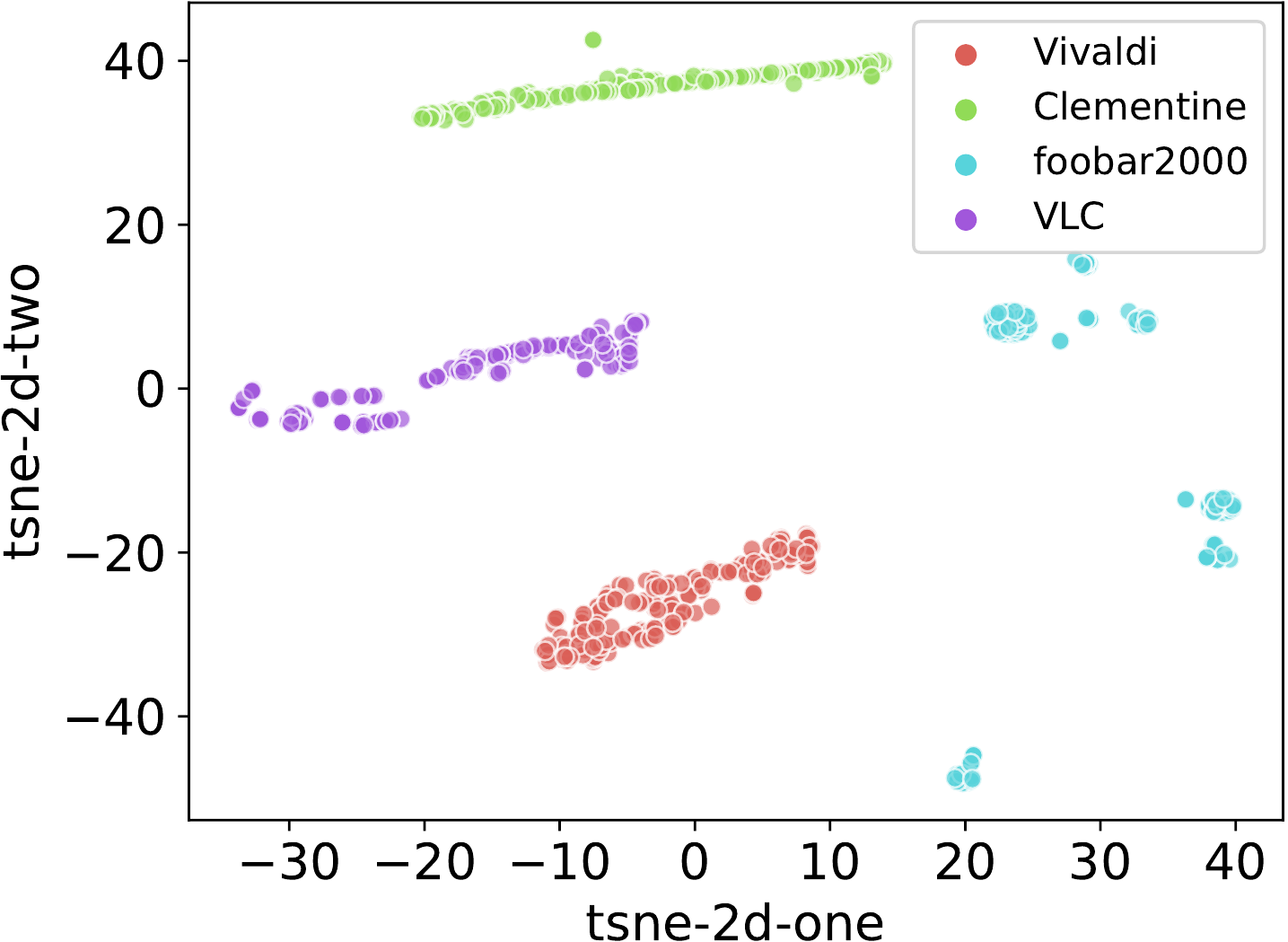}}
\caption{Two-dimensional scatter plots of various clusters using
  t-SNE.} \label{fig:instance_validation}
\vspace{-2ex}
\end{figure}

With $18$ programs, there are $12240 = {18 \choose 3} \times 15$ experimental instances that
could be examined. We carried out the experiments with $5000$ random
instances, resulting in a detection ratio of $98.7\%$. In
Table~\ref{tab:clustertesting}, we present $15$ such experimental
instances, among which the proposed approach fails to detect the
unknown program in two instances. Analyzing the data of these two
instances, we observed that performance counter data of Vivaldi and
Brave (i.e., the instance in row $6$) are very close. Similar trends were
also observed in the case of row $9$ between Firefox and Internet
Explorer. In reason of this behavior could be the fact that all the
web browsers use the same rendering engine. In
Fig.~\ref{fig:instance_validation}, we present several two-dimensional
scatter plots of data points randomly selected from $4$ programs using
t-SNE. We observe that there is a clear separation among clusters. We
verified the cluster index obtained by the clustering algorithm
against ground truth data. To determine the value of parameter $\beta$
(i.e., the distance threshold of deviation), we performed a grid
search in the range $[0, 1]$ and selected the value that yields the
best performance.      

\begin{table}[t]
\centering
\caption{Selected Instances for Testing of Clustering Model}\label{tab:clustertesting}
\begin{tabular}{|c|l|l|c|}
\hline
\multirow{2}{*}{\bf No.} &  \multicolumn{2}{c|}{\bf Experimental Instances} & \multirow{2}{*}{\bf Result} \\ 
\cline{2-3}
& Seeding Clusters & Test & \\
\hline
1&gVim, Notepad, Notepad++ &  Atom& Detected \\
\hline
2& Atom, Brave, Google Chrome  & gVim & Detected\\
\hline
\multirow{2}{*}{3}& Sublime Text, Google Chrome,  & \multirow{2}{*}{Notepad} & \multirow{2}{*}{Detected} \\
& Vivaldi && \\
\hline
4&Atom, Firefox, Vivaldi & Notepad++ &  Detected \\
\hline
\multirow{2}{*}{5}&Brave, Google Chrome,  & Sublime  &  \multirow{2}{*}{Detected} \\
& Vivaldi & Text & \\
\hline
\multirow{2}{*}{6} & \multirow{2}{*}{gVim, Vivaldi, Clementine} & \multirow{2}{*}{Brave} &  Not  \\
&&& Detected\\
\hline
\multirow{2}{*}{7}&Notepad, Internet Explorer,  & Google  &  \multirow{2}{*}{Detected} \\
& WildBit & Chrome & \\
\hline
8&Atom, Vivaldi, Clementine & Firefox &  Detected \\
\hline
\multirow{2}{*}{9} & \multirow{2}{*}{gVim, Firefox, VLC} & Internet  &  Not  \\
&&Explorer& Detected\\
\hline
\multirow{2}{*}{10} & Notepad++, FastStone,  & \multirow{2}{*}{Vivaldi} &  \multirow{2}{*}{Detected} \\
& foobar2000 && \\
\hline
11&Atom, Brave, Clementine &  WildBit &  Detected \\
\hline
\multirow{2}{*}{12}&Sublime Text, Brave, & \multirow{2}{*}{FastStone} &  \multirow{2}{*}{Detected} \\
& Google Chrome && \\
\hline
13&Atom, Notepad, Vivaldi &  ImageGlass &  Detected \\
\hline
\multirow{2}{*}{14} & Clementine, Google Chrome,  &  \multirow{2}{*}{MS Paint} &  \multirow{2}{*}{Detected} \\
& Internet Explorer & & \\
\hline
15&Atom, Brave, Vivaldi & XnView &  Detected \\
\hline
\end{tabular}
\vspace{-1.5ex}
\end{table}

\begin{figure}[t]
\centering
\subfigure[Web browser samples at the coarse-grained  level.]{\includegraphics[width=0.40\textwidth]{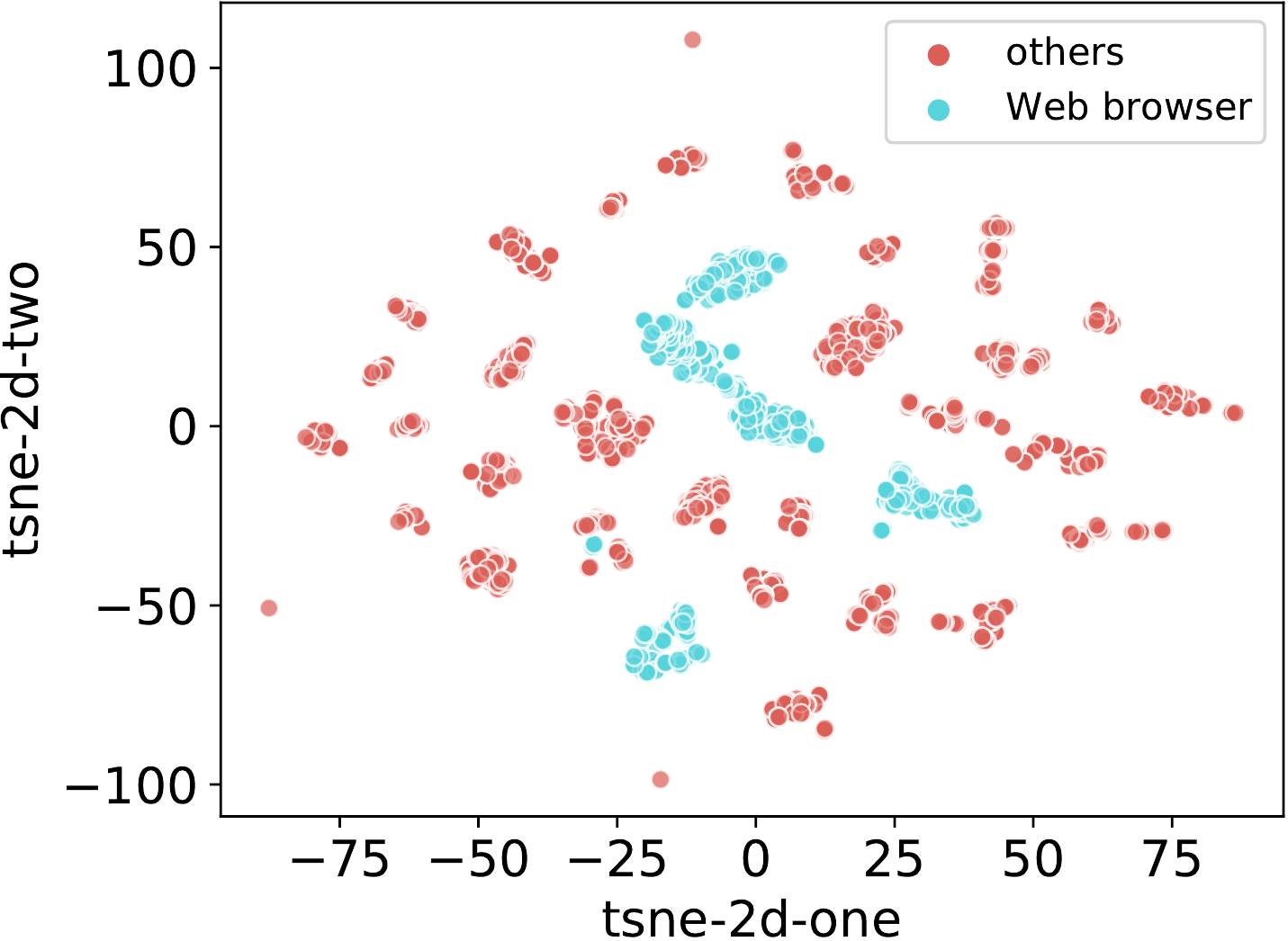}}
\subfigure[ Firefox samples at the fine-grained
level.]{\includegraphics[width=0.40\textwidth]{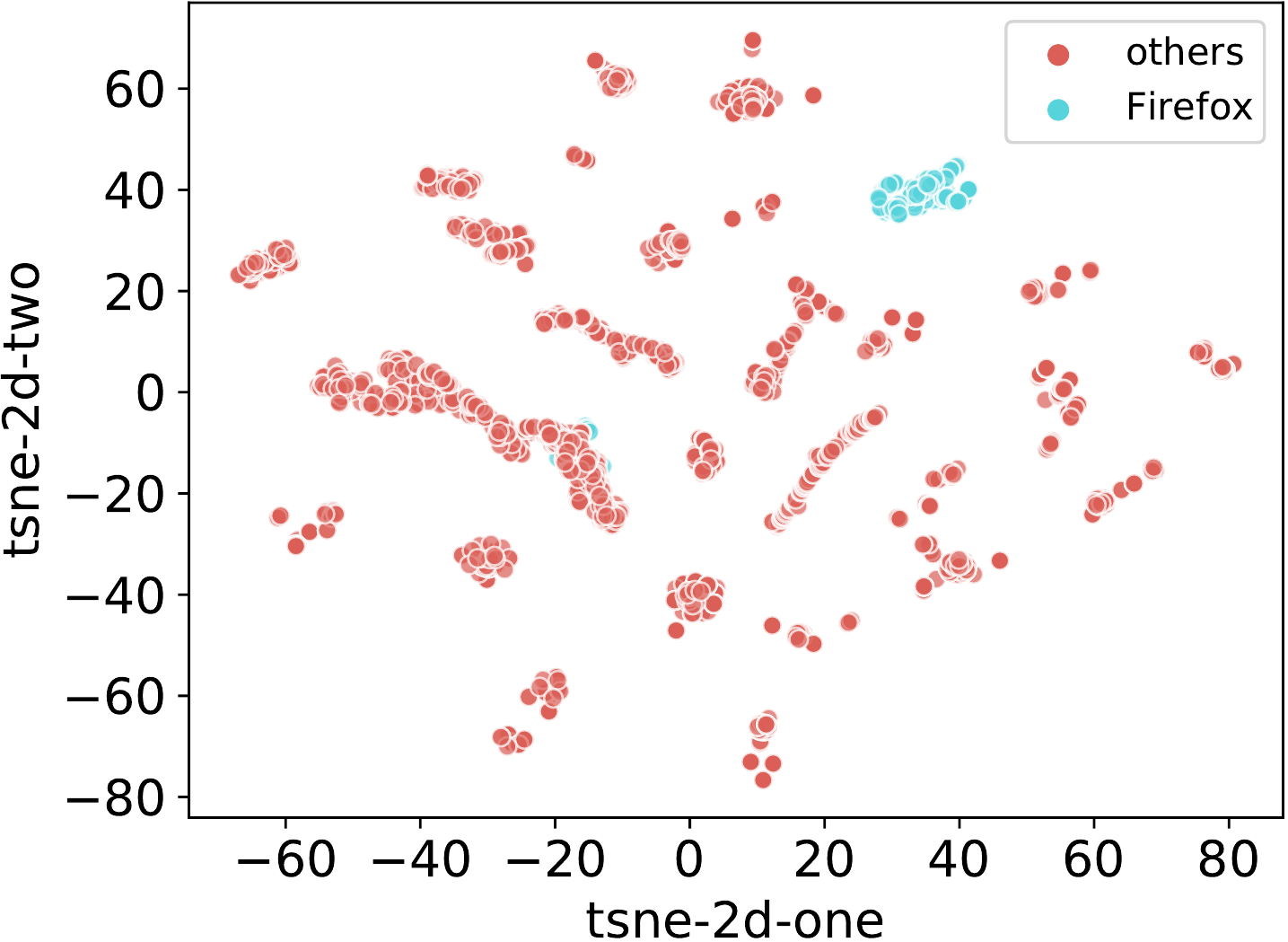}}
\vspace{-2ex}
\caption{t-SNE visualization.} \label{fig:3_1_17_1}
\vspace{-3ex}
\end{figure}

We also performed experiments of the detection of unknown programs at
the coarse-grained and fine-grained levels. At the coarse-grained
level, we considered all the seeding programs are clustered into $3$
clusters while the unknown program should form the fourth cluster. At
the fine-grained level, there will be $17$ seeding clusters and the
unknown program should form the $18^\text{th}$ cluster. In
Fig.~\ref{fig:3_1_17_1}, we present the t-SNE plots for these two
experiments with labels obtained from our clustering algorithm. The
results show that in both cases, the unknown program forms a new
cluster that is well separated from the seeding clusters.

\subsubsection{Detection of Behavior Deviation}
\label{subsec:res_dev_det}

\begin{figure}[!b]
\centering
\subfigure[Page File Bytes]{\includegraphics[width=0.23\textwidth]{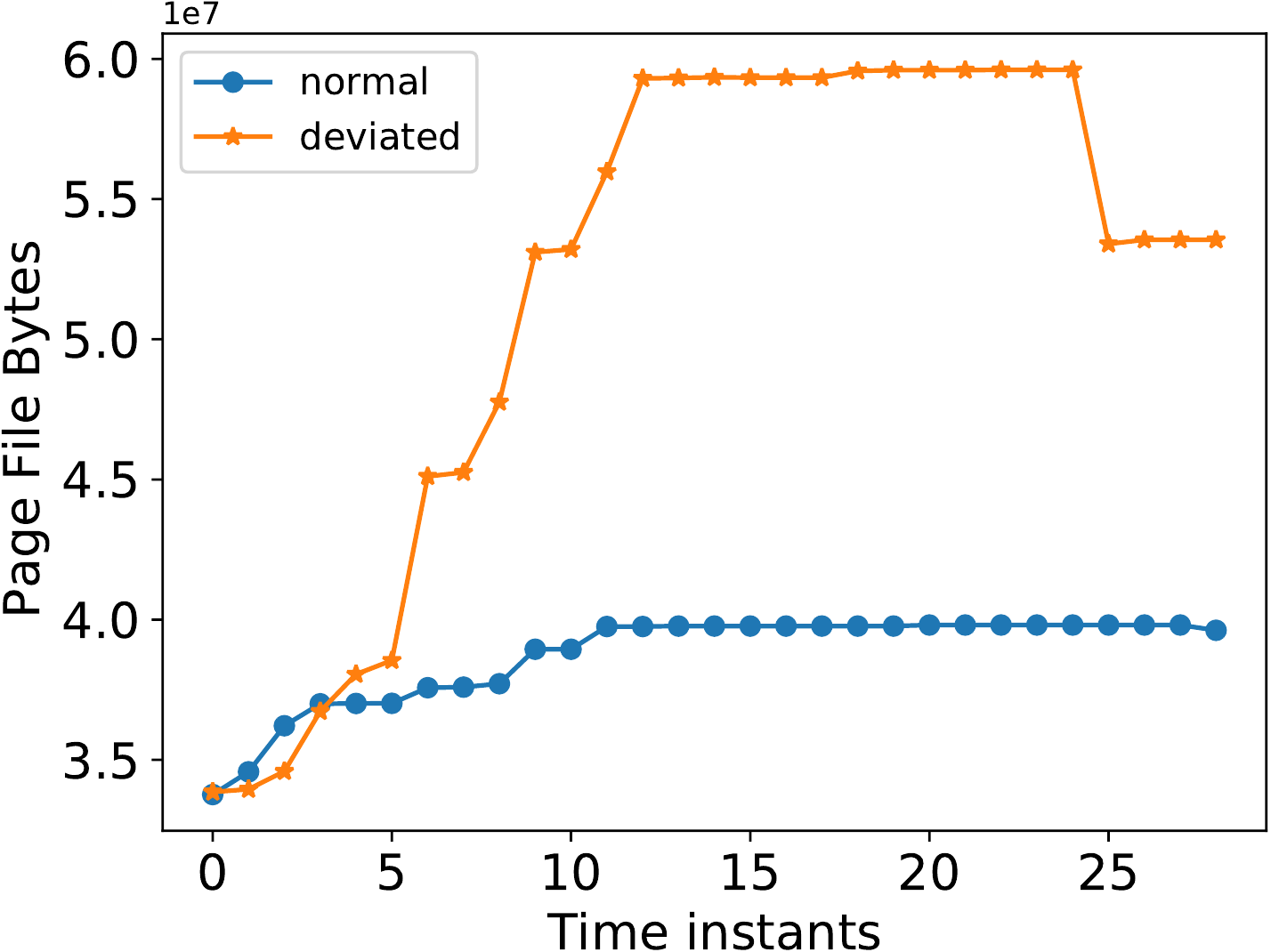}}
\subfigure[I/O Data Operations]{\includegraphics[width=0.23\textwidth]{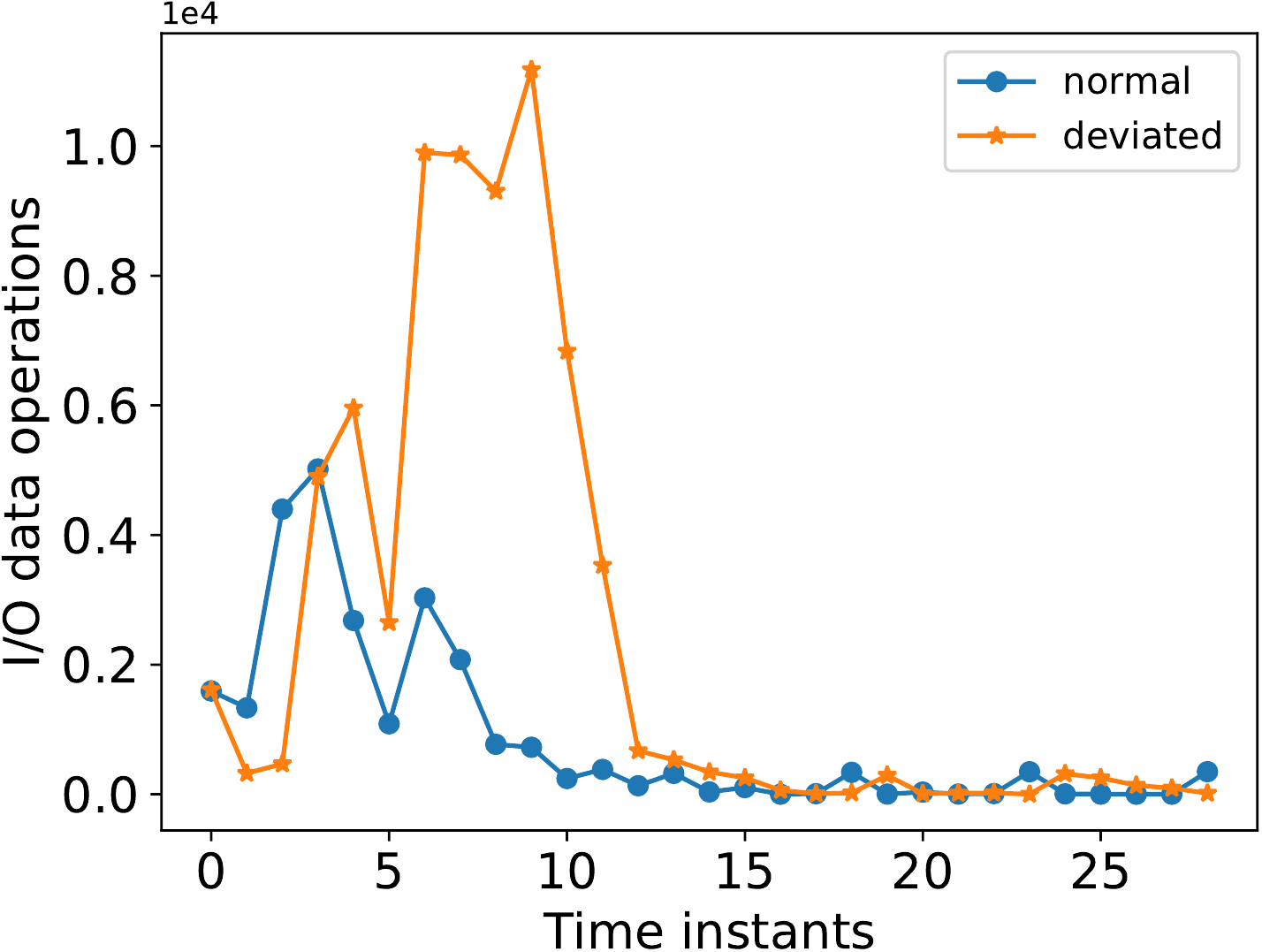}}
\subfigure[Handle Counts]{\includegraphics[width=0.23\textwidth]{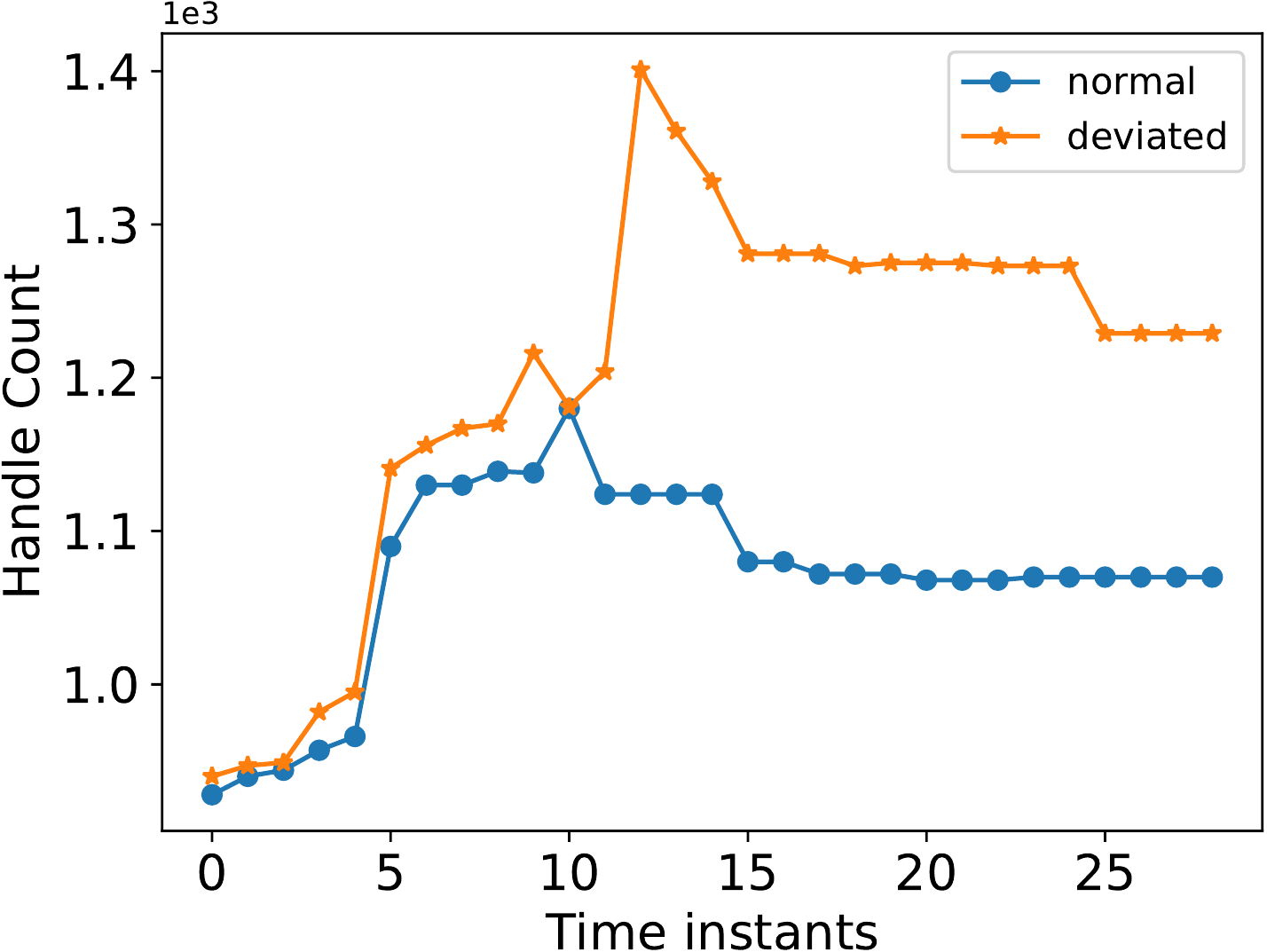}}
\subfigure[Pool Paged Bytes]{\includegraphics[width=0.23\textwidth]{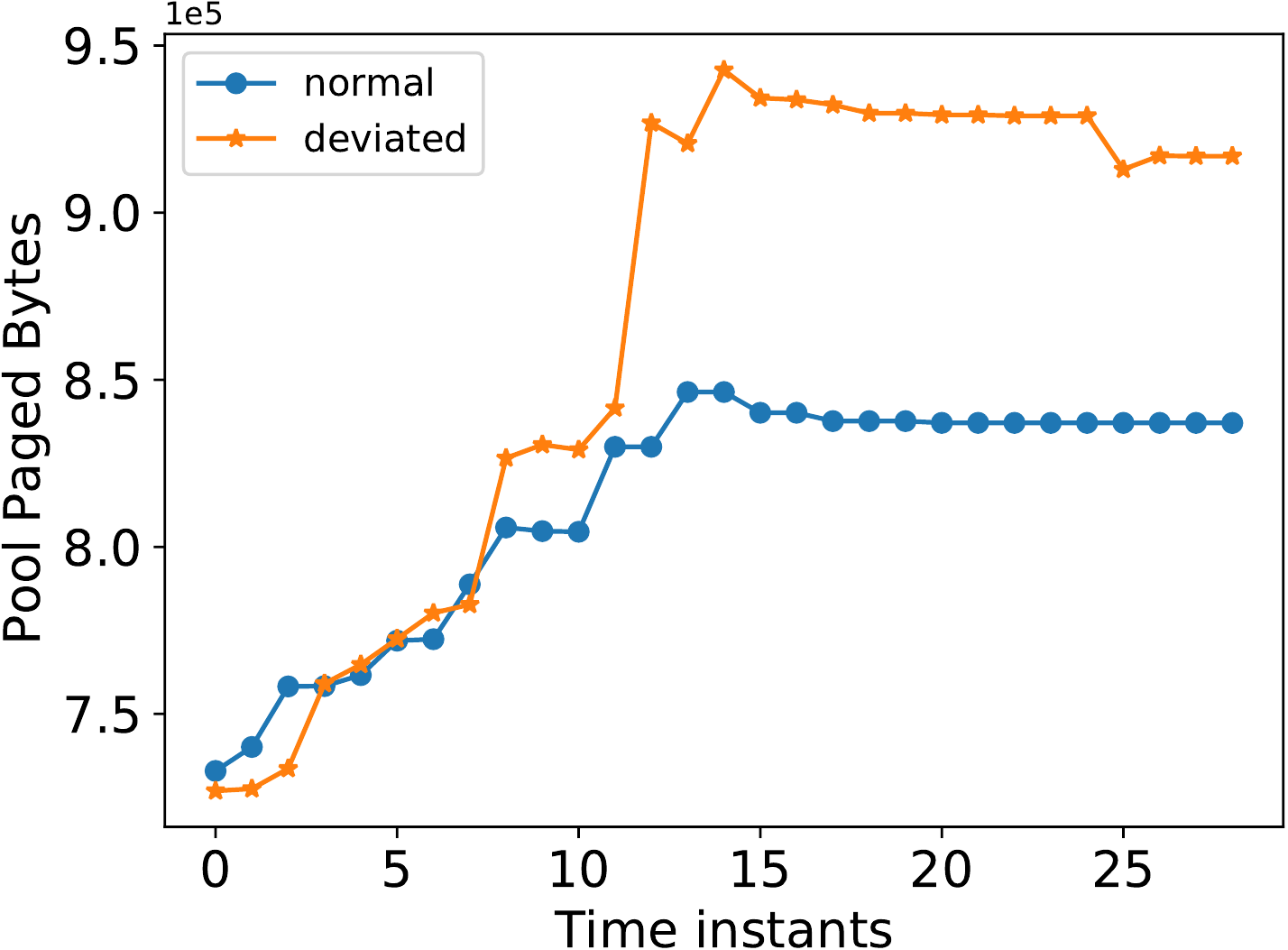}}
\caption{Various performance counter time series for normal and
  deviated versions of Google
  Chrome.}
\label{fig:dev_vs_undev_many_counters}
\vspace{-2.5ex}
\end{figure}

As we discussed earlier, existing programs (that have been installed
in the systems) could change their normal behavior due to various
reasons such as updates or being tampered by viruses or
attackers. While program updates may not cause any harmful
consequences, programs tampered by viruses or attackers may lead to
data loss. Thus, it is beneficial to detect such deviation as earlier
as possible. In this experiment, we demonstrate that the proposed
approach could detect such a behavior deviation. We introduced a
deviation in Google Chrome with a custom extension, which
consecutively opens many photos on the Amazon website. In
Fig. \ref{fig:dev_vs_undev_many_counters}, we present the deviation of
performance counter data (i.e., \texttt{Page File Bytes}, \texttt{I/O Data Operations},
\texttt{Handle Counts} and \texttt{Pool Paged Bytes}) in comparison with one of the data
samples collected during its normal working conditions, i.e., the data
samples collected with Google Chrome when opening the Amazon website without the customized extension. We used Algorithm~\ref{alg:clustering} to determine whether a data sample of a program has ``significantly'' deviated from its normal behavior. We have also run the original $k$-means clustering algorithm on the dataset of Google Chrome that included the data samples collected during normal working conditions and the samples that are collected with the custom extension. The clustering results show that the data samples are separately clustered into two groups. As the label of the samples collected during normal conditions is known, the remaining cluster includes all the data samples collected with custom extension, representing the behavior deviation of Google Chrome.

\subsubsection{Bare-metal vs Virtual Environment}

\begin{figure}[t]
    \centering
    \subfigure[Handle Counts Time Series.]{\includegraphics[width=0.40\textwidth]{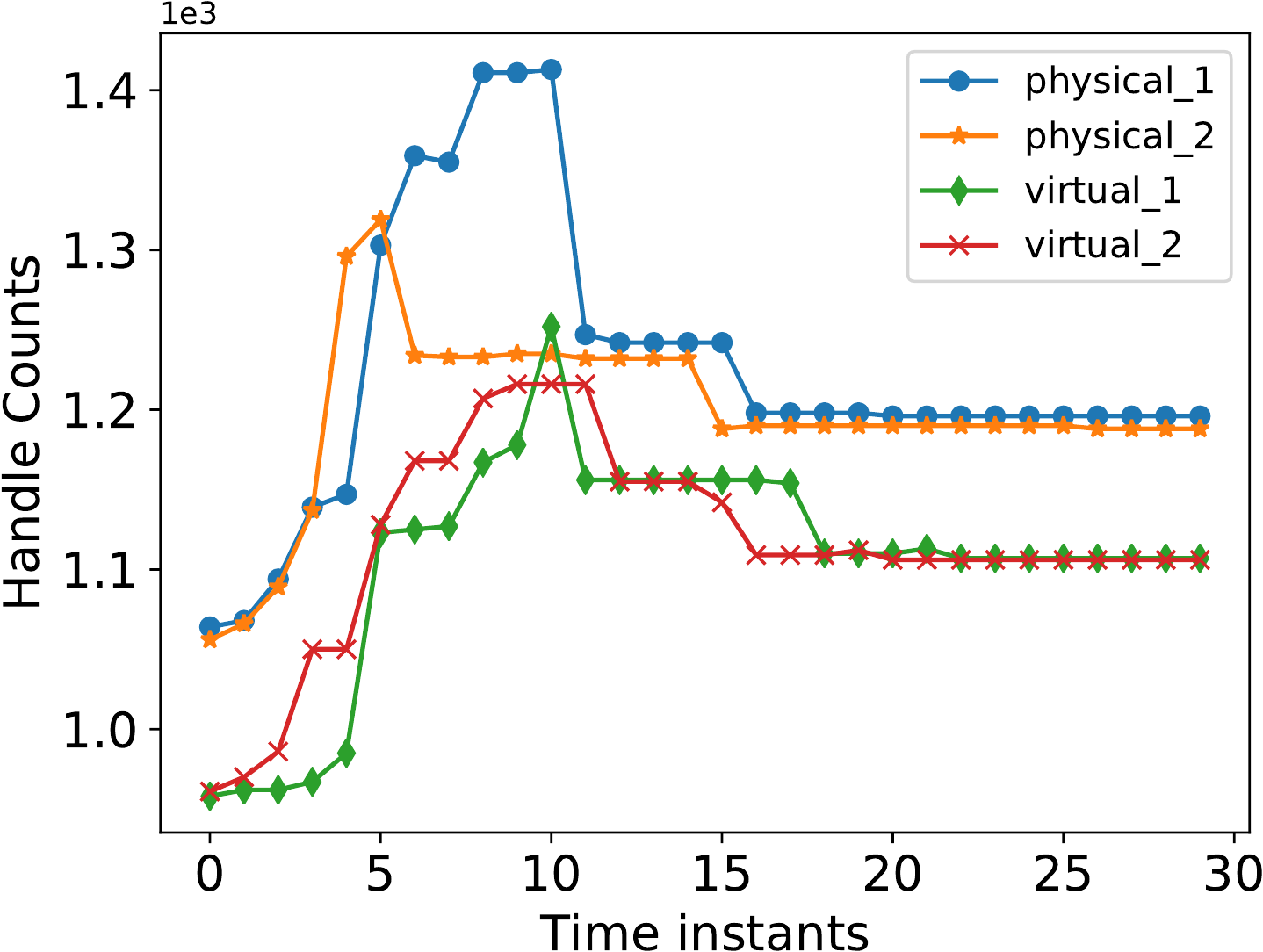}}
    \subfigure[Page File Bytes.]{\includegraphics[width=0.40\textwidth]{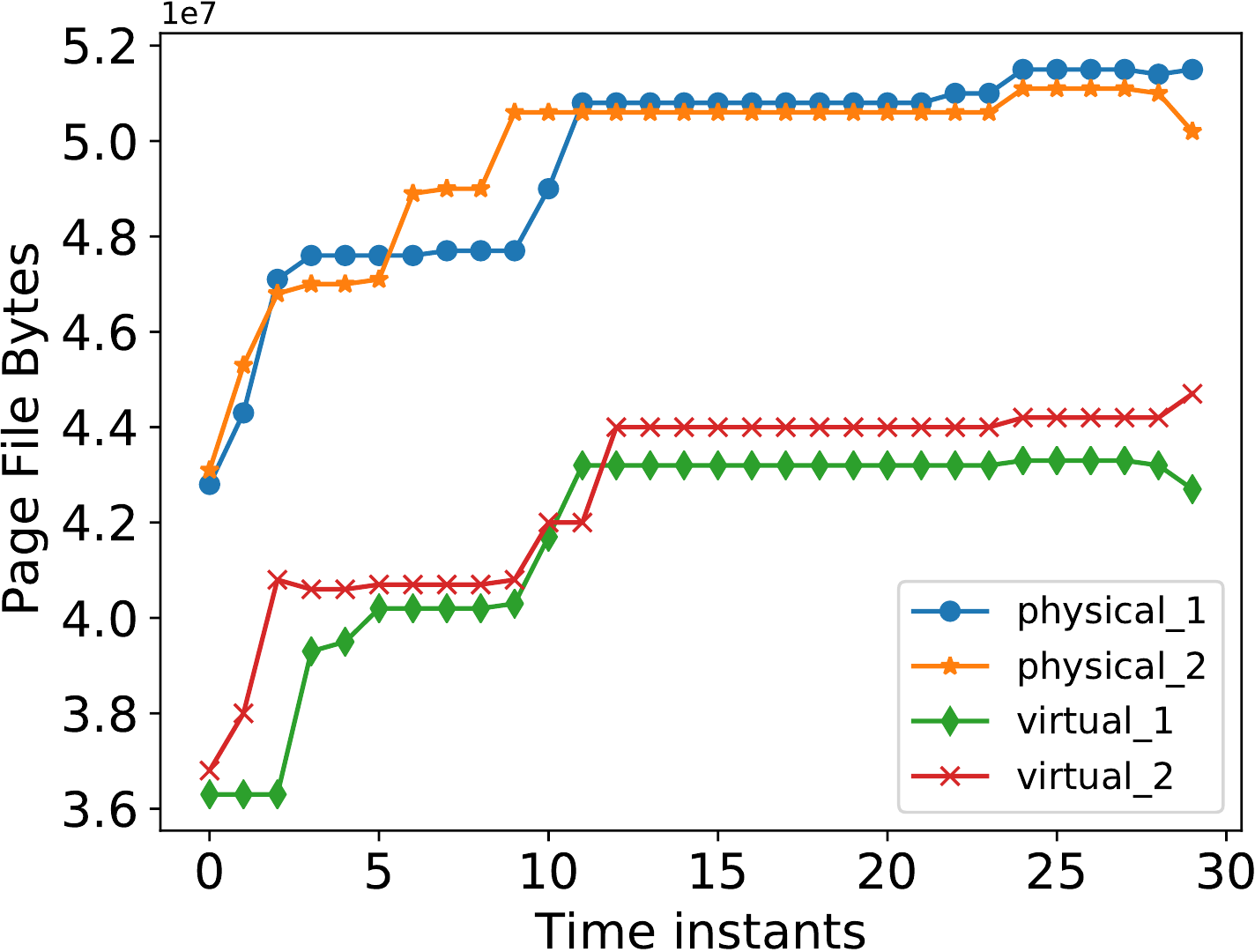}}
    \caption{Difference in performance counter values in virtual and
      physical environment for two separate
      runs.} \label{fig:physical_vs_virtual}
    \vspace{-2ex}
  \end{figure}
  
While we have done all the experiments in a virtual environment (i.e.,
executing the programs in a virtual machine and collect performance
counter time series for analysis), the proposed approach can also be
applied to a bare-metal environment. To demonstrate this, we have run
the \texttt{perfextract} tool in a physical host and collected the
performance counter time series of the programs. In
Fig.~\ref{fig:physical_vs_virtual}, we plot the time series of two
performance counters (\texttt{Handle Counts} and \texttt{Page File
  Bytes}) of Google Chrome when running in a virtual machine and in a
bare-metal environment. We can observe that there is a change in the
values of the performance counters (i.e., the number of events of
\texttt{Handle Counts} and \texttt{Page File Bytes} in the virtual
machine is fewer than that in a bare-metal environment). However, evolution trends remain the same for both environments. As we discussed
earlier, the proposed approach learns not only the dynamic
behavior of a program from the values of performance counters but also
the correlation among the values through its changing trends (i.e.,
evolution trends) over time. Thus, the clustering model when deployed
to the bare metal environment may need to be retrained on the data
collected in this environment so as to adapt to the new data distribution
and achieve the desired performance.

\section{Conclusions}\label{sec:conclusion}

In this paper, we presented a clustering-based approach to analyze
program behavior in a given environment using performance counters. We
collected the data from various performance counters as time
series and analyzed program behavior based on temporal and spatial
correlations represented by the time series. We developed a tool to
ease the collection of data using existing primitives provided by the
operation systems (e.g., Windows OS provides \texttt{TypePerf} to
collect the current value of a performance counter). We adopted
a conventional clustering algorithm, $k$-means clustering, to cluster
programs at both coarse-grained level (i.e., based on type of
programs) and fine-grained level (i.e., differentiating among
programs). We also adopted the algorithm to detect new/unknown
programs installed in the system, as well as behavior deviation of
existing programs due to software updates or tampering activities. We
carried out experiments with $18$ programs that belong to $4$
different groups (web browsers, text editors, image viewers and audio
players). The experimental results show that the proposed approach
manages to accurately cluster programs into their respective
groups. The results also demonstrate that the proposed approach can be
used to detect whether a new program emerges (i.e., it is a
newly-installed program) or an existing program has deviated from its
normal behavior. For future work, we aim at further developing the
proposed approach to perform malware detection using performance
counter time series.

\begin{acks}
This research is supported by the Agency for Science, Technology and
Research (A*STAR) under its RIE2020 AME Core Funds (SERC Grant
No. A1916g2047). We would like to thank Partha Pratim Kundu, Sin
G. Teo and Vasudha Ramnath for their valuable feedback and fruitful
discussion during the implementation of this work. 
\end{acks}

\balance
\bibliographystyle{ACM-Reference-Format}
\bibliography{dynamics2020}
\end{document}